\documentclass[traditabstract]{aa} 

\usepackage{txfonts} 
\usepackage{natbib}
\bibpunct{(}{)}{;}{a}{}{,} 

\usepackage{graphicx}

\newcommand \ba{{ BATSE}}
\newcommand \fe{{\it Fermi}}
\newcommand \ep{$E_{\rm peak}$}
\newcommand \epo{$E^{\rm obs}_{\rm peak}$}
\newcommand \al{$\alpha$}
\newcommand \epop{$E^{\rm obs}_{\rm peak}$--$P$}
\newcommand \epof{$E^{\rm obs}_{\rm peak}$--$F$}
\newcommand \liso{$L_{\rm p,iso}$}
\newcommand \eiso{$E_{\rm iso}$}
\newcommand \gsim{ \lower .75ex \hbox{$\sim$} \llap{\raise .27ex \hbox{$>$}} } 
\newcommand \lsim{ \lower .75ex\hbox{$\sim$} \llap{\raise .27ex \hbox{$<$}} } 

\newcommand \ama{$E_{\rm peak}-E_{\rm iso}$}
\newcommand \yone{$E_{\rm peak}-L_{\rm iso}$}

\newcommand \F{$F$\ }
\newcommand \Pf{$P$\ }

\begin{document}
\title{Spectral properties of long and short Gamma--Ray Bursts: 
comparison between BATSE and \fe\ bursts}

\author{L. Nava \inst{1,2}\thanks{lara.nava@sissa.it}
 \and G. Ghirlanda \inst{2} \and G. Ghisellini \inst{2} \and A. Celotti \inst{1}}
\institute{SISSA, via Beirut 2-4, I--34151 Trieste, Italy
\and Osservatorio Astronomico di Brera, via E. Bianchi 46, I--23807 Merate, Italy }
\date{Received .. ... .. / Accepted .. ... ..} 

\abstract{
We compare the spectral properties of 227 Gamma Ray Bursts (GRBs) 
detected by the \fe\ Gamma Ray Burst Monitor (GBM) up to February 2010 
with those of bursts detected by the CGRO/BATSE instrument. 
Out of 227 \fe\ GRBs, 166 have a measured peak energy \epo\ of their 
$\nu F_{\nu}$ spectrum: of these 146 and 20 belong the long and short class, respectively.
\fe\ long bursts follow the correlations defined by BATSE bursts 
between their  \epo\  vs fluence and peak flux: as already shown for the latter ones, 
these correlations and their slopes do not originate from instrumental selection effects.
\fe/GBM bursts extend such correlations toward lower fluence/peak 
energy values with respect to BATSE ones whereas no GBM long burst 
with  \epo\ exceeding a few MeV is found,  despite the possibility of detecting them.
Again as for BATSE, $\sim$ 5 \% of long and almost all short GRBs detected by 
\fe/GBM  are outliers of  the \ep--isotropic equivalent energy (''Amati'') correlation 
while no outlier (neither long nor short) of the \ep--isotropic equivalent luminosity (``Yonetoku") correlation 
is found. 
\fe\ long bursts have similar typical values of  \epo\ but a harder 
low energy spectral index with respect to all BATSE events, 
exacerbating the inconsistency with the limiting slopes of 
the simplest synchrotron emission models.  
Although the short GRBs detected by \fe\ are still only a few, we confirm that  
their  \epo\ is greater and the low energy spectrum is harder than those of long ones. 
We discuss the robustness of these results with respect to observational biases induced by 
the differences between the GBM and BATSE instruments.
} 

\maketitle

\section{Introduction}

Our current knowledge of the spectral properties of the prompt emission in GRBs 
mainly relies on the data collected in almost 10 years by the Burst And Transient Source 
Experiment (BATSE) onboard the {\it Compton Gamma--Ray Observatory (CGRO)}
satellite. 

BATSE was composed by eight identical modules each containing two NaI(Tl) scintillation detectors: a Large 
Area Detector (LAD), optimised for sensitivity and directional response,  and a 
Spectroscopy Detector (SD) optimised for energy band and resolution. 
Thanks to its wide energy coverage ($\sim$25 keV - $\sim$1 MeV) and all--sky 
viewing, BATSE detected more than 2700 GRBs. The published spectral catalogs comprise
relatively small sub-samples of bright GRBs, selected on the 
basis of the burst fluence and/or peak flux \citep{preece00,kaneko06}. 

The analysis of such samples revealed two main results about the spectral and energetics properties
of these GRBs. 

The first one is that the prompt spectra can be typically 
well described by a curved function showing a peak - in a $\nu F_\nu$ representation - at a typical 
energy \epo\ of a few hundreds of keV but whose distribution 
spans nearly three orders of magnitude. Large dispersions characterise also the distributions of the low--
and high--energy photon indices, whose characteristic values are
$\alpha\sim -1$ and $\beta\sim -2.3$, respectively \citep{preece00,kaneko06}.

The second one is that the rest frame peak energy \ep\ of long bursts correlates 
both with the bolometric isotropic energy \eiso\ emitted during the prompt 
\citep{amati02} and with the bolometric isotropic luminosity \liso\ estimated at 
the peak of the light curve \citep{yonetoku04}. Such correlations are clearly
an intriguing powerful clue about the dominant emission mechanism 
during the prompt phase. Furthermore, as by correcting by the presumed jet opening 
angle the dispersion of the correlations even reduces \citep{ghirlanda04a}, 
GRBs become potential standard candles \citep{ghirlanda04b}. 

\cite{nava08} (N08 hereafter) performed the spectral analysis on a larger fainter sample of BATSE bursts,
selected by setting a limiting fluence to $F=10^{-6}$ erg/cm$^2$. They found that the 
spectral parameter \epo\ shows a strong correlation with the fluence $F$ (or the peak flux $P$). 
These imply that the derived distribution of \epo\ is strongly affected by a selection based on cuts 
in fluence (or peak flux).
The correlations in the observer frames (\epo--\F and \epo--\Pf)
may be just the consequence of the rest frame correlations mentioned above. Alternatively, it has 
been claimed that the rest frame correlations are the result of instrumental 
selection effects \citep{band05,nakar05}. 

\citet[hereafter G08]{ghirlanda08} and N08 examined two instrumental selection effects 
which may affect the relation between \epof\ and \epop\ 
and found that - although they do affect the burst sample properties - 
they are not responsible for the correlations found in the observational planes.   
Moreover \citet{ghirlanda10} recently showed that the correlation \ep-- \eiso\  and 
\ep--\liso\ holds for the time--resolved quantities within individual \fe\ bursts (see also \citealt{firmani09} 
for Swift bursts and \citealt{krimm09} for Swift--Suzaku GRBs) and that this 
``time--resolved" correlation is similar to that defined by the time integrated properties 
of different GRBs. These results support the hypothesis that the "Amati" and "Yonetoku" correlations 
do have a physical origin and that the trends seen in the "observational planes" are just 
their outcome.

The same analysis has been applied to short bursts \citep{ghirlanda09}, though their lower fluence 
limits the possibility to determine with reasonable accuracy the
properties of a statistically significant sample of events. Also in this case 
spectral information mainly arise from BATSE data.

Evidence of a spectral diversity between long and short bursts comes from their different
hardness--ratios (HR) \citep{kouveliotou93}.
The larger HR of short bursts might be ascribed to a larger \epo. 
However, due to the relation between \epo\ 
and the bolometric fluence and peak flux, a direct comparison between the \epo\ distributions 
of the two different burst classes must take the different fluence/peak flux selection criteria into account.  
\citealt{ghirlanda09} (G09 hereafter) performed a detailed spectral analysis of  
samples of short and long BATSE bursts selected on the same limit on the peak flux.
They found that the peak energy distributions of the two classes are similar, while 
the most significant difference concerns the low--energy power--law indices, with short bursts having
typically a harder $\alpha\sim -0.4$. 

The existence of intrinsic spectral--energy correlations for short bursts is still an 
open question, due to the paucity of short GRBs 
with measured redshift. 
However, the findings by G09 show that short events do not follow the \ep--\eiso\ 
correlation defined by long ones, while their are consistent with a similar 
\ep--\liso\ correlation \citep{amati06,ohno08,krimm09}.

The \fe\ satellite, launched in June 2008, opened a new and powerful
opportunity to shed light on the origin of the 
GRB prompt emission thanks to its two high energy instruments: the Large Area Telescope 
(LAT) and the Gamma--ray Burst Monitor (GBM). LAT detected in 1.5 year 
very high energy emission ($>$ 100 MeV) from 15 GRBs, out of which 12 are discussed 
in \citet{ghisellini10}.

The GBM extends the BATSE energy range at both low and high 
energies. Its twelve NaI detectors assure a good spectral resolution between $\sim$8 keV and 
$\sim$1 MeV and two BGO detectors extend the instrument sensitivity up to $\sim$40 MeV. 
While - as for BATSE - the NaI detectors guarantee full--sky coverage, the smaller geometric area (16 times lower than that
of the LADs) result in a lower sensitivity. 

Until now (beginning of March 2010), about 280 triggers of the GBM are reliably 
classified as GRBs\footnote{http://heasarc.gsfc.nasa.gov/W3Browse/fermi/fermigbrst.html}. While a spectral catalogue has not yet been published, 
preliminary results of the spectral analysis performed by the GBM team have been 
distributed to the community through the Galactic Coordinates Network (GCNs) Circulars.
For 166 events the spectral properties (\epo) are well constrained. 

Their statistically significant number allows to start a meaningful comparison between BATSE and GBM results.
The fact that the two instruments have a different energy range and effective area sensitivities
could be in turn exploited to infer the properties of GRBs prompt emission. This is the aim of 
this work.  In other words we try to assess whether 
the GBM is confirming any/which ones of the BATSE results 
or is revealing something new about prompt spectra in the hard X/$\gamma$--ray energy range
and the existence of the above mentioned spectra vs energetics correlations. 

In particular the GBM wider energy window should allow to examine two important aspects:
on one hand the low energy photon index - as first inferred from BATSE data - indicate 
(\citealt{preece98a,preece02,ghisellini00}) an inadequacy of 
a simple synchrotron model to account for the GRB prompt emission; on the other hand 
better constraints on 
the high energy spectral index may help to discriminate between quasi--thermal 
and non--thermal 
processes and would give clues on the energy distribution of the emitting electrons. 
Moreover, due to its wider energy range, the GBM could more easily detect 
soft and hard bursts, i.e. those events whose peak energy is outside the BATSE energy range. 

In this paper we study these issues with \fe/GBM bursts.
In \S 2 we present the samples of \fe\ bursts (both long and short) 
and their spectral parameters, as reported in the GCNs. 
In \S 3 the properties of \fe/GBM bursts are compared with the BATSE ones in terms of
\epo, \F\ and \Pf, i.e. in the \epof\ and \epop\ planes.
We compute the relevant instrumental selection effects introduced  
by the GBM characteristics on these quantities, considering separately short and long GRBs. 
We compare the spectral parameters of BATSE 
and \fe\ bursts in \S 4 and discuss in \S 5 the similarities and differences of the spectral 
properties of GRBs 
between the two instruments and, for each of them, 
between the population of short and long events. 

\section{Samples and spectral models}
\label{models}
The spectral analysis performed by different authors \citep[K06]{preece00,sakamoto05,butler07,nava08} revealed that
time--integrated and time--resolved burst spectra are (mostly) best fitted by different models, the simplest ones being: 
i) a single power law model (PL), ii) a Band function (BAND), 
which consists of two smoothly connected power laws and iii) a Comptonized model (COMP), 
i.e. a power law with a high energy exponential cutoff. 
In some cases a smoothly broken power law with a flexible curvature (SBPL) better fits the 
observed spectra (K06). 

A simple PL function clearly indicates that no break/peak energy is detected within the energy range of the instrument.
However, it is also statistically the best choice 
when the signal to noise is very low (since this model has the lowest number of free 
parameters), as is the case for most of Swift/BAT bursts. 

The BAND model \citep{band93} has four parameters to describe the low 
and high power law behaviours, the spectral break and the flux normalisation. Typically, 
the low energy photon index  $\alpha >  -2$ [$N(E)\propto E^{\alpha}$] 
and the high energy photon index $\beta < -2$ [$N(E)\propto E^{\beta}$],
so that a peak in $\nu F_\nu$ can be defined, with \epo\ 
ranging between a few keV and a few MeV. 

When there is no evidence for a high energy photon tail or \epo\ 
is near the high energy boundary of the instrument sensitivity (and $\beta$ is poorly 
constrained) a COMP model is preferred, due to the lower number of parameters. 
Also in this case a peak energy can be defined when $\alpha>-2$. 
In the BAND model the spectral curvature is fixed by $\alpha$, 
$\beta$ and \epo. Sometimes it is necessary to introduce a fifth parameter that 
defines its sharpness, to account for very sharp breaks 
up to the limit of a broken power law function. 
Analytical expressions and a more 
detailed description of these different spectral models can be found in K06.

\subsection{Long bursts}

For the BATSE instrument we will refer to two different samples of 
long bursts with published spectral properties: the bright bursts sample analysed by 
K06 and the sample of 100 fainter bursts published in N08. 

The K06 sample\footnote{http://heasarc.gsfc.nasa.gov/W3Browse/cgro/batsegrbsp.html} 
contains {\it all} BATSE bursts with peak flux  
$P~(50-300 \rm~keV)>10~photons~cm^{-2}~s^{-1}$ {\it or} fluence 
$F~(\sim20-2000\rm~keV)>2\times10^{-5} \rm~ergs~cm^{-2}$.  
From the total 350 events we select the long ones, i.e. with observed 
duration $T_{90}>2$ s. Since we are interested in comparing the distribution
of \epo, we exclude bursts whose best fit model is a PL 
or for which the best fit is a curved model, but with $\alpha<-2$ or 
$\beta>-2$. This selection results in 280 GRBs: for 104 events 
the spectrum is best fitted by a BAND model, for 65 by a COMP model and for 
111 by a SBPL model.

The K06 analysis enlarges previous spectral samples based on BATSE data 
\citep{preece00} but still contains only the brightest objects. 
For this reason N08 selected a sample of 100 fainter bursts, representative 
of the $\sim1000$ BATSE bursts with fluence
$10^{-6} <$ $F$ $<2\times10^{-5}\rm~erg~cm^{-2}$. Of these 
44 bursts are best described by the COMP model, 44 by the BAND model, and 
12, not showing evidence of curvature, by a PL function and thus are excluded.

A spectral catalogue of \fe\ bursts has not yet been published. 
However, results of preliminary spectral analysis  performed by the 
\fe/GBM team can be found in the GCN archive. We collected 
all bursts 
with spectral information up to the beginning of March 2010 (227 objects) and
selected the 146 of them with known \epo\ and belonging to the long class, 
without any cut in fluence or in peak flux.
According to the results reported in the GCN, the COMP 
model better fits the spectrum in 90 cases, while in 56 cases a BAND model 
is required. While clearly the analysis reported in the CGN might be in some 
cases preliminary, we expect this does not affect the statistical results of our
analysis.

\subsection{Short bursts}

The most comprehensive sample of short BATSE GRBs
with well defined spectral parameters is composed by the 79 events analysed by \citet{ghirlanda09},
selected for having $P >3~\rm photons~cm^{-2}~s^{-1}$. 
In most cases their spectrum is best fit with a COMP model. 

For the \fe/GBM instrument we still refer to the spectral results reported 
in the GCN archive. Among the 227 GBM bursts with 
known \epo\ 20 are short (9 are fitted with a COMP model and 11 with a BAND model). 

\section{\epo--Fluence and \epo--Peak Flux planes: comparison between BATSE and GBM bursts}
\label{piani osservativi}

\begin{figure*} 
\includegraphics[scale=0.56]{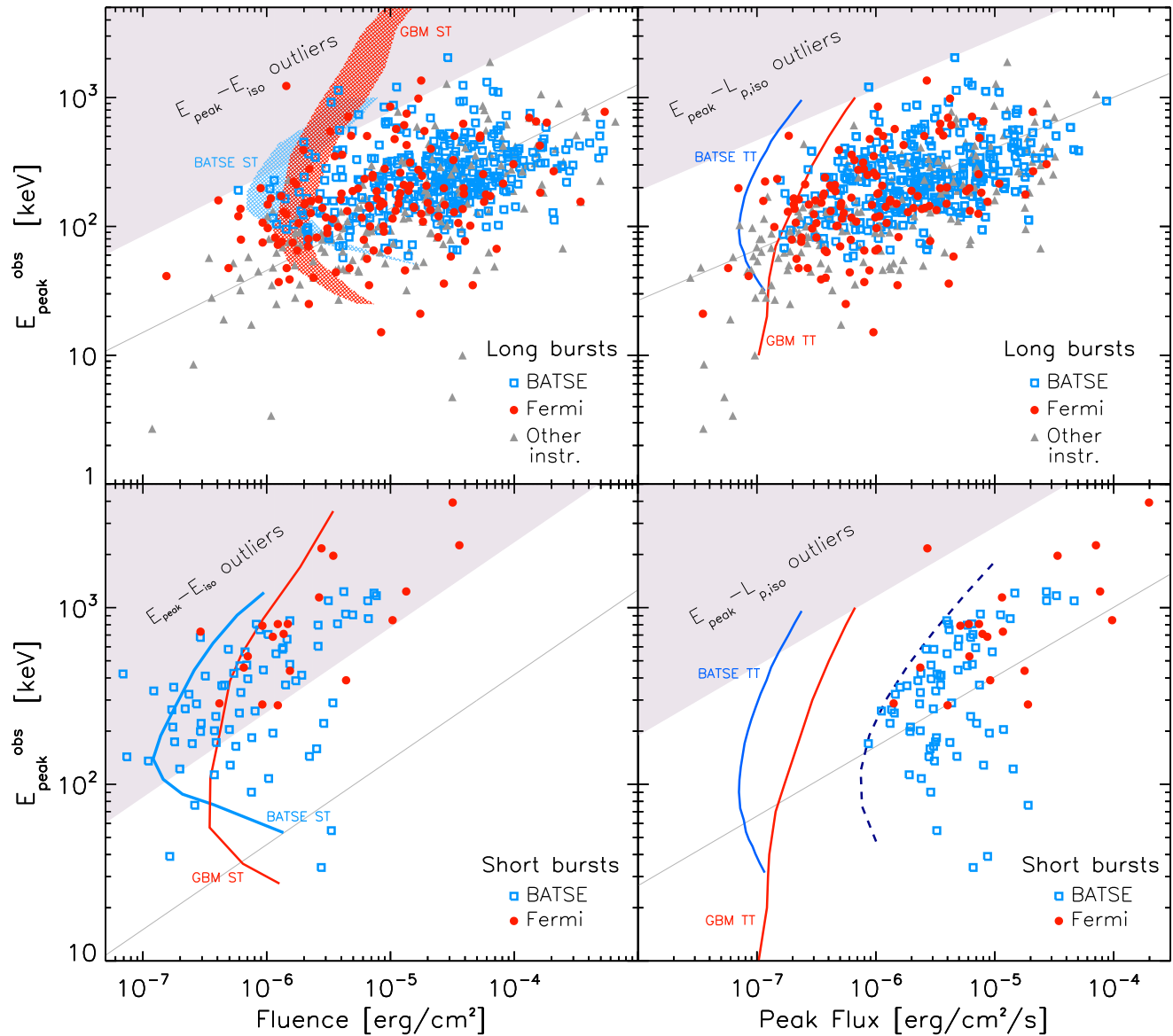}
\caption{
\epo--Fluence and \epo--Peak Flux planes for long (upper panels) 
and short (bottom panels) bursts. Empty squares represent BATSE bursts, 
filled circles \fe/GBM bursts and filled triangles indicate  
events detected by other instruments (from N08). 
In all panels the instrumental 
limits for BATSE and \fe/GMB are reported: shaded curved regions in the 
upper left panel show the ST, estimated for bursts lasting 5 and 20 s; 
solid curves in the bottom left panel represents the ST for short bursts. 
Solid curves in the right panels define the TT, identical for short 
and long events. 
Thresholds for BATSE are taken from G08 while those for the GBM instrument 
are derived in this work.
The dashed curve in the bottom right panel represents the selection 
criterion applied by G09 for their sample of short bursts, i.e. $P>$ 3 phot/cm$^2$/s.} 
\label{piani correle} 
\end{figure*} 

The distribution of GRBs with and without measured redshift in the planes 
\epo--\F\ and \epo--\Pf\ has been investigated by G08 and N08 with 
the main purpose of probing the role of instrumental selection effects on the 
two rest frame corresponding correlations, i.e. the  \ep--\eiso\  
(the ``Amati" correlation - \citealt{amati02}) and \ep--\liso\ 
(the ``Yonetoku" correlation - \citealt{yonetoku04}). 
A correlation between the total fluence and \epo\ 
was found by Lloyd, Petrosian \& Mallozzi (2000) for a sample of \ba\  
bursts without measured redshifts.  
This finding has been recently confirmed 
by \citet{sakamoto08} using a sample of bursts detected by Swift, BATSE and Hete--II. 
In particular, they note that X--Ray Flashes and X--Ray Rich bursts 
satisfy and extend this correlation to lower fluences. 
N08 considered all events with published spectral 
information detected by different instruments (Swift, BATSE, Hete--II, 
Konus/Wind and {\it Beppo}Sax) together with the 100 fainter BATSE bursts.
In both planes long bursts define a correlation, with fainter bursts having lower \epo.
 
In order to examine the distribution of \fe\ bursts in the 
observational planes \epo--\F\ and \epo--\Pf\ and compare it with that 
of BATSE events it is 
mandatory to estimate the biased induced by instrumental selection effects
(see G08). 
One bias is represented by the capability of an instrument 
to be triggered by a burst, i.e. the ``trigger threshold" (TT). A second one 
concerns the limit imposed on the photon number in order to significantly model the spectrum, 
defined ``spectral threshold" (ST) by G08 and N08.
The TT translates into a minimum peak flux, which depends on the burst spectrum and in particular on \epo\
and can be described as a curve in the \epo--\Pf\ plane.
The second requirement (ST) results into a minimum fluence which not only depends on  
\epo\ but also on the burst duration. 
For this reason the ST is represented as a region in the 
\epo--\F\ plane.

Another relevant point is to assess whether \fe\ bursts 
are consistent with following the rest--frame correlations defined by BATSE GRBs.
In the observational planes it is possible to define a ``region of outliers" 
(\citealt{nakar05}; G08; N08) for the corresponding rest--frame correlations 
(i.e. the ``Amati" and the ``Yonetoku" for the \epof\ and \epop\ planes 
respectively): bursts lying in this region, no matter their redshift, 
cannot be consistent with correlations. 

N08 find that 6\% of BATSE long bursts are outliers of the 
\ep--\eiso\ correlation defined by the sub--sample of bursts with known 
redshift, while no outlier is found for the \ep--\liso\ correlation. 

For short bursts the situation is quite different and has been 
investigated by G09. The peak flux selection criterion of their sample clearly 
determines the shape of the distribution of data points at low peak fluxes. 
This fact hamper the possibility of determining the existence of a correlation, 
even though a weak indication for a trend is present (high fluences events, 
not affected by selection effects, tend to have high \epo\ values). 

Interesting results instead regard outliers: almost all short bursts 
are outliers of the \ep--\eiso\ correlation defined by long bursts
but they can be consistent with their very same \ep--\liso\ 
correlation. These findings are supported by the properties of 
the very few short bursts with measured redshift with respect to the 
rest--frame correlations (\citealt{amati06}; G09).

\subsection{Estimate of GBM instrumental selection effects}

Following G08, the TT curves are obtained adapting the results of Band 
(2006) and are shown in the right (top and bottom) panels of Fig. \ref{piani correle}. 
These are clearly the same for long and short bursts as the trigger 
threshold depends only on the peak flux. 

The ST curves have been calculated from numerical simulations, as done 
by G08 for BATSE and {\it Swift}/BAT. 
To perform these simulations, the typical background spectrum and the 
detector response function must be known.
For the bursts detected by the GBM both of them depend on 
several factors (e.g. the satellite attitude when a burst occurs), and thus 
there is no universal background and response matrix which can be adopted. 
To overcome this problem we use the real 
backgrounds and responses of several GRBs detected by the GBM and 
average the results of the simulation to build average ST curves 
for the population of long and short bursts, respectively.

To be consistent with the spectral analysis of the GBM data reported in the GCNs, 
our simulation performs a joint spectral analysis of the two most illuminated 
NaI detectors and the most illuminated BGO detector. For each detector we 
extract the background spectrum and consider the response files. 
Simulations were performed for the 10 long bursts published in \citet{ghirlanda10}
for two representative values of the duration, $T_{90} =$ 5 
and 20 s (corresponding in Fig. 1 to the curves 
delimiting the red shaded region at low and high fluence, respectively).
For short bursts we estimated the ST curves for all of the 20 short bursts of the 
GBM sample assuming a typical duration of the simulated spectra of 0.7 s (red curve in the bottom left panel of Fig. \ref{piani correle}). 
This value, also adopted by G08, is  the typical duration of short GRBs observed by the GBM. 

For the TT and ST curves of the BATSE instrument we simply report the 
results published in  G08 (for long bursts) and in G09 (for short bursts).

\subsubsection{Long bursts}

\emph{\epo\ vs Fluence:}  as can be seen in the top left panel of Fig. \ref{piani correle},
the distribution of GBM bursts in this plane (filled circles) 
displays some differences compared with that of BATSE bursts (empty squares).

The presence of \fe\ bursts with low \epo\ (between $\sim$10 and $\sim$50 keV), not present in the 
BATSE sample, is clearly due to the wider energy range of the GBM 
instrument, sensitive down to $\sim$8 keV (see the ST curves). In this region, GBM bursts are consistent 
with bursts detected by other instruments (filled triangles). 
Also  GBM bursts define a correlation which mostly overlaps with that 
defined by BATSE bursts and extends to the lower-left part of the  \epo--\F\ plane.

Despite the GBM assures good coverage up to $\sim$30 MeV (vs $\sim$ 1 MeV of BATSE),
the distribution of \epo\ at high energies is similar for the two instruments. 
This could suggest that long bursts with very high \epo\ (grater than 1 MeV) 
do not exist or are very few. Indeed, we note that high \epo\ also means
GRBs with high fluences, which are rarer than GRBs with lower fluences.
In fact, the fluence distribution of GBM bursts (Fig. \ref{fluence_tutti}, 
empty histogram) is peaked at $F \sim 7.8\times10^{-6}$ erg cm$^{-2}$, which 
corresponds to typical \epo$\gtrsim$100 keV. 

Note that while BATSE bursts appear concentrated at higher \epo--\F, 
the sample is composed by \emph{all} bursts with 
$F>2\times10^{-5}$ erg cm$^{-2}$ and only \emph{one hundred} of fainter 
bursts representative of $\sim1000$ objects with 
$10^{-6}$ erg cm$^{-2} < F < 2\times10^{-5}$ erg cm$^{-2}$. 
The fluence distributions of these two different samples 
are reported in Fig. \ref{fluence_tutti}. The insert shows 
the whole fluence distribution for BATSE bursts, built up weighing 
the fainter class by a factor 10. We refer to this "weighted sample" as "BATSE Total". 
The comparison of the fluence distribution of the BATSE Total sample 
with the GBM one (solid blue and orange lines in the 
insert of Fig. \ref{fluence_tutti}) reveals that they are 
similar (KS probability $P=0.06$, see Table \ref{tab uno}). 

Another result shown by the GBM bursts and consistent with the conclusion 
drawn from BATSE is that the ST are not responsible for the particular 
distribution of the data in the plane. Although they 
prevent us from detecting bursts with very low fluence, they 
cannot explain why bursts tend to distribute along a correlation. 
This is well visible for BATSE bursts: events with large \epo\ tend to 
concentrate far from the ST, i.e., at higher fluences. The trend 
shown by the BATSE ST (which requires higher limiting fluences when 
\epo\ is very high and very low) cannot explain this behaviour. 
The same holds for GBM bursts, for which the ST are even less curved 
and cannot be responsible for the observed 
correlation which has a slope $\sim0.20\pm0.04$, consistent with 
that defined by BATSE bursts (N08).

The sensitivity of the GBM instrument also allows to probe the region of 
outliers at high \epo. Consistently with BATSE, we found that about 
the 5\% of \fe/GBM bursts are outliers with respect to the 
"Amati" correlation defined by the sample of GRBs with redshift 
(see \citealt{ghirlanda10} for a recent update). 

\emph{\epo\ vs Peak Flux:}
the distribution of GBM bursts with respect to BATSE bursts is shown in the upper right panel of Fig. \ref{piani correle}. 
Solid curves represent the TT derived for both instruments 
(adapted from \citealt{band06}). 
On average, the GBM instrument is a factor 3 less sensitive than BATSE
in the common energy range. 
As remarked by N08, the sample of BATSE bursts lies far from its TT, 
suggesting that for this instrument the 
demand of performing a reliable spectral analysis dominates the selection. 
This is not the case for the GBM: the data points lie very near 
the TT curves, suggesting that if a burst is detected there is a 
good chance to recover its spectral shape with a certain precision. 
This makes TT and ST competitive selection effects for \fe/GBM bursts. 
As for BATSE, none of the GBM bursts lies in the region of outliers: 
they can be all consistent with the Yonetoku correlation. 

The \epo\ distribution is shown in Fig.~\ref{ep_tutti} for both instruments.
The central values of the distributions, obtained by modelling them
with a gaussian function, are reported in Table \ref{tab uno}. 
The BATSE bright bursts analysed by K06 have a \epo\ distribution 
peaked around 250 keV, at considerably higher energies 
with respect to that of GBM bursts which peaks at 140 keV 
(KS probability $P=4\times10^{-15}$). The difference strongly reduces 
when considering the BATSE Total sample, including both bright (K06) and faint 
(N08) events. 
The insert in Fig. \ref{ep_tutti} shows their  \epo\ 
distribution compared with that of \fe/GBM bursts.  
They are very similar, both in terms of central value and width of the 
gaussian fit ({see Table \ref{tab uno}), suggesting that the two instruments 
are seeing the same burst population (KS probability $P=0.01$). 
The only difference regards the low energy end of the distributions.
For BATSE bursts it has a quite 
sharp cutoff at $\sim$ 50 keV (green hatched histogram in Fig. \ref{ep_tutti}), according 
to simulations performed by G08 (see
Fig. \ref{piani correle}) showing that it is very difficult for 
this instrument to recover \epo\ $<$ 50 keV. Large fluences, unusual for such values of 
\epo, would be required. 
For \fe\ the situation is different: the GBM is sensitive down to 
$\sim$ 8 keV and this should introduce a cutoff on the \epo\ distribution 
at lower values. A sharp cutoff, however, is not visible for two reasons: i) the 
smoother cut introduced by the TT and ST for the GBM instrument compared to that
for the BATSE instrument (upper panels in Fig. \ref{piani correle}); ii) the \fe\ sample is not selected in fluence 
(a fluence limit at $10^{-6}$ ergs cm$^{-2}$ would deplete the lowest bins, producing 
a less symmetric distribution). 

\begin{table}
\begin{tabular}{c|ccc|c||c}
\hline \hline
Parameter             &        &BATSE   &         &\fe\   &KS    \\
                      &Bright  &Faint   &Total    &        &  \\
\hline
Fluence [erg/cm$^2$]  &3.5e-5  &4.9e-6  &5.7e-6   &7.8e-6  &0.06   \\
\epo\ [keV]           &  251   &141     &158      &144     &0.01   \\
\hline
\hline
\end{tabular}
\caption{Central values of the \epo\ and fluence distributions reported 
in Fig. \ref{fluence_tutti} and Fig. \ref{ep_tutti}. The last column 
reports the KS probability that the distributions for the complete sample of 
BATSE bursts (BATSE Total, see text) and for the \fe/GBM bursts are drawn 
from the same parent population.}
\label{tab uno}
\end{table}

\subsubsection{Short bursts}
For short bursts (bottom panels in Fig. \ref{piani correle}) the situation is 
quite different in both planes. 

Also for the GBM there is only a weak indication of a correlation in both planes, consistent with the trends 
suggested by BATSE short bursts. 
However, the overall behavior is different from what happens for long GRBs.

At high \epo\ GBM short events occupy the same region of the BATSE ones, and even 
extend the trend \epo-\F\ to \epo\ above the BATSE energy range, 
revealing that \epo\ larger than 1 MeV are common in short bursts.

Furthermore, contrary to what one might expect, they appear missing below \epo$\sim 250$ keV. 
At low fluence this can be accounted for by considering the ST derived for the GBM instrument: indeed
Fig. \ref{piani correle} shows that GBM short events lie very near to the ST curve (bottom left panel of Fig. \ref{piani correle}), that prevents the estimate of \epo $< 250$ keV when $F<5\times10^{-7}$erg cm$^{-2}$.

The higher sensitivity of BATSE instead 
implies that its ST is located at lower fluence. 
This however does not account for the absence of short GBM burst at \epo$< 250$ keV and higher fluence ($> 5\times10^{-7}$ erg cm$^{-2}$):
the BATSE sample shows that GRBs with low \epo\ and on the right side of the GBM ST do exist. 
Their absence in the present GBM sample seems to imply that they are relatively rare and therefore that at a fluence $>5\times 10^{-7}$ erg cm$^{-2}$ most of the events
should have \epo$>$ 250 keV. In other words this would support the presence of a \epo-\F\ (broad) correlation. 
A larger sample of short GBM bursts may however reveal some of them. 

In terms of \epo\ this translates into a distribution peaked at higher energies 
compared both to the BATSE one and to the GBM long events.

In the \epo--\Pf\ plane similar conclusions can be drawn: for both 
instruments the TT curves (solid lines) do not affect the samples.
The BATSE one is clearly limited by the selecting cut applied on the peak 
flux (shaded curve), while we remind that GBM bursts have not been selected a priori  
either in fluence or in peak flux and their presence requires their detection and the
possibility of performing a significant spectral analysis.  
From the fact that the peak flux of \fe\ GRBs is significantly above their TT, we infer 
that their selection is dominated by the ST.

Almost all \fe\ short bursts lie in the \epo--\F\ region of outliers and are  
inconsistent at more than 3$\sigma$ with the Amati relation 
defined by the sample of 105 long bursts (including also 10 \fe\ bursts). 
This confirms what found with short BATSE bursts (G09). 

On the contrary, only one burst appears to be an outliers of the \ep--\liso\ correlation 
of long GRBs. 
The hypothesis that short and long bursts follow the same \ep--\liso\ 
correlation is supported by the few short events with known redshift.
\begin{figure} 
\vskip -0.3 cm
\includegraphics[scale=0.55]{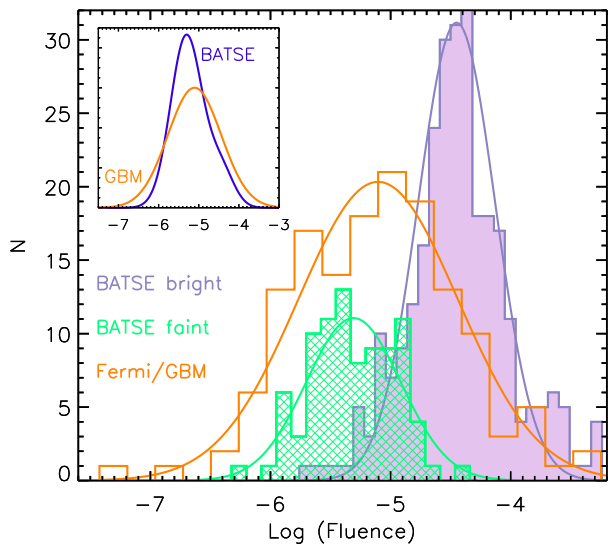}
\caption{
Fluence distributions for BATSE bright bursts (filled 
histogram, from K06), BATSE faint bursts (shaded histogram, from N08) 
and \fe/GBM bursts (empty histogram). 
The insert compares the fluence distribution for the BATSE Total sample (see text) with that for \fe/GBM bursts.} 
\label{fluence_tutti} 
\end{figure} 
\begin{figure} 
\vskip -0.3 cm
\includegraphics[scale=0.55]{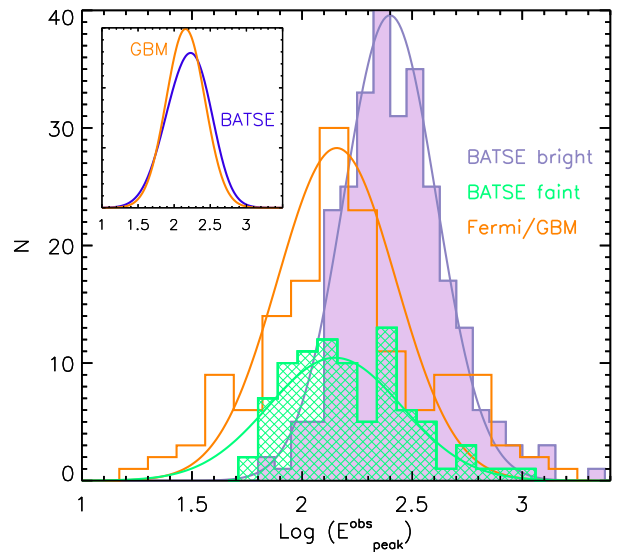}
\caption{
\epo\ distributions for BATSE bright bursts (filled histogram, from K06), 
BATSE faint bursts (shaded histogram, from N08) and 
\fe/GBM bursts (empty histogram). 
The insert compares the \epo\ distribution for the BATSE Total sample (see text) with that for \fe/GBM bursts.} 
\label{ep_tutti} 
\end{figure} 

\section{Long bursts: parameter distributions and model biases}

Spectral parameters, such as \epo\ and $\alpha$, show different
typical values and width of their distributions depending on the fitting
spectral model (K06). 
For example, $\alpha$ (\epo) 
derived from the COMP model is on average softer (harder) than 
that derived for bursts whose best fit model is a BAND function; 
the low energy power--law index distribution for the SBPL 
is softer than that derived from the COMP and the BAND models. 
These differences can be ascribed both to biases introduced by 
models with fewer parameter or by 
real intrinsic differences between spectra best described by 
different functional forms.

In this section we quantify them for the BATSE sample
and then investigate if similar behaviours are found in the \fe/GBM one. 
We consider only long bursts, as the sample of short GBM bursts is 
still scarce.

\begin{table*}
\begin{center}
\begin{tabular}{c|ccc|cccc|cc}
\hline
\hline
Model &         &         BATSE         &        &          &\fe/GBM &           &   &  KS&(GBM vs BATSE) \\
      &$\alpha$ &$\alpha_{\rm eff,25}$   & \epo\   &$\alpha$ &$\alpha_{\rm eff,8}$   &$\alpha_{\rm eff,25}$ &\epo\  & $\alpha$     & \epo\    \\
\hline
COMP   &--1.14   &--1.22  &  311   &--0.94  &--0.98  &--1.10  &  145  & 0.004  &  $1\times10^{-9} $\\
BAND    &--0.93   &--1.04  &  232   &--0.76  &--0.80  &--0.89  &  143  & 0.04    &  $1.5\times10^{-5}$ \\  
SBPL     &--1.25   &           &  232   &            &           &            &          &            &    \\ 
BAND + COMP   &--1.00   &--1.12  &  264   &--0.88  &--0.92  &--1.05  &  144  & 0.005  &   $5\times10^{-12}$   \\        
\hline
ALL       &--1.11   &--1.16  &  251   &--0.88   &--0.92 &--1.05   &  144  & $6\times10^{-9}$&   $4\times10^{-15}$   \\
\hline
\hline
\end{tabular}
\caption{Central values of the distributions of $\alpha$, $\alpha_{\rm eff}$ and \epo. 
The last two columns give the KS probability resulting from the comparison 
of the BATSE and GBM bursts distributions of $\alpha$ and \epo. 
}
\label{modelli}
\end{center}
\end{table*}

\subsection{\epo\ distribution}
\begin{figure} 
\includegraphics[scale=0.55]{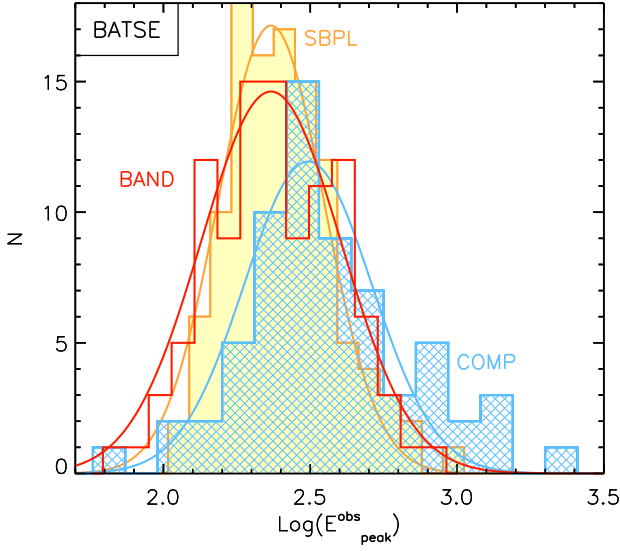}
\caption{
\epo\ distributions for different spectral models (bright BATSE bursts). 
The three histograms refers to events whose spectrum is best fitted 
by a SBPL function (filled), COMP model (shaded) and BAND one (empty).} 
\label{ep_k06} 
\end{figure} 

At first, we consider the sample of bright BATSE bursts (K06) 
according to their best fit model (COMP, BAND or SBPL). 

Fig.~\ref{ep_k06} shows the distribution of \epo\ for each model. 
The BAND and SBPL models provides very similar central values 
($\langle E^{\rm obs}_{\rm peak, SBPL}\rangle\simeq\langle 
E^{\rm obs}_{\rm peak, BAND}\rangle\simeq 232$ keV;  KS probability $P=0.366$), 
while higher  \epo\  are derived with the COMP model 
($\langle E^{\rm obs}_{\rm peak, COMP}\rangle\simeq 311$ keV). 
Consistently with the results by K06, the KS probability between the distributions from the COMP and 
BAND models is $P=3\times10^{-4}$, while it is $P=1.5\times10^{-4}$
between the COMP and SBPL models.

Two different factors can account for the higher \epo\ of spectra fitted with the COMP model.  
When \epo\ is close to the high energy bound of the instrument 
$\beta$ can be hardly constrained: in these cases a COMP model 
describes the data equally well, but it is statistically preferred as it has a lower 
number of free parameters.
Moreover, the simulations by K06 show that the COMP model tends to systematically 
overestimate the best fit \epo\ with respect to the simulated value and thus
underestimates the hardness of the low energy spectrum. 
These effects can account for the different \epo\ and $\alpha$ distributions 
between the COMP and BAND models found here (Fig. \ref{ep_k06} and Fig. \ref{a_fermi} 
--upper panel-- respectively).  

\begin{figure} 
\includegraphics[scale=0.55]{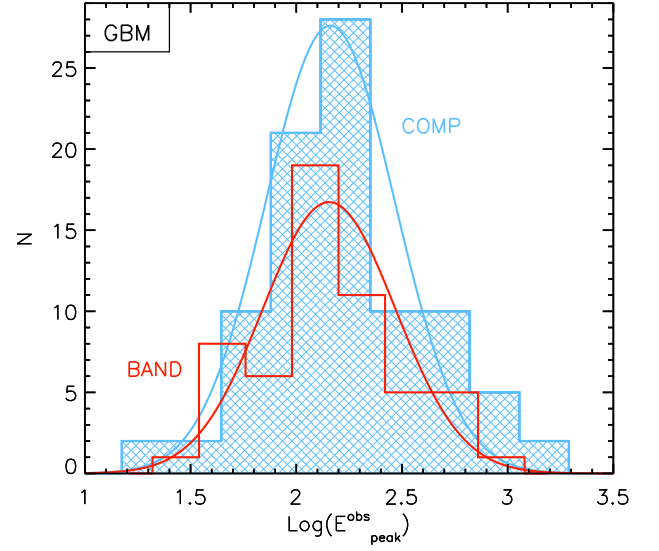}
\caption{
\epo\ distributions for different spectral models 
(\fe/GBM bursts). 
The shaded (empty) histogram refers to bursts whose spectrum is 
best fitted by a COMP (BAND) model.} 
\label{ep_fermi} 
\end{figure} 

Fig. \ref{ep_fermi} shows the \epo\ distribution for \fe/GBM bursts, 
according to the best fit model (empty and filled histograms for the BAND and COMP models, respectively). 
The two distributions appear quite similar 
(KS probability = 0.553), both in terms of central values 
(see Table \ref{modelli}) and widths. 
Contrary to what found with BATSE data, there is no indication of a bias 
introduced by the COMP model.

The KS test indicates a significant  
inconsistency between the \epo\ distributions of bright BATSE and GBM bursts (last column 
in Table \ref{modelli}). However, 
when faint bursts (N08) are included, the BATSE Total 
and GBM sample distributions are totally consistent,
(see Table \ref{tab uno} and the insert in Fig. \ref{ep_tutti}), 
indicating that the two instruments 
are seeing the same population of bursts in terms of \epo$\sim 
150$ keV. 

\subsection{$\alpha$ distribution}
\label{alpha}

\begin{figure} 
\includegraphics[scale=0.6]{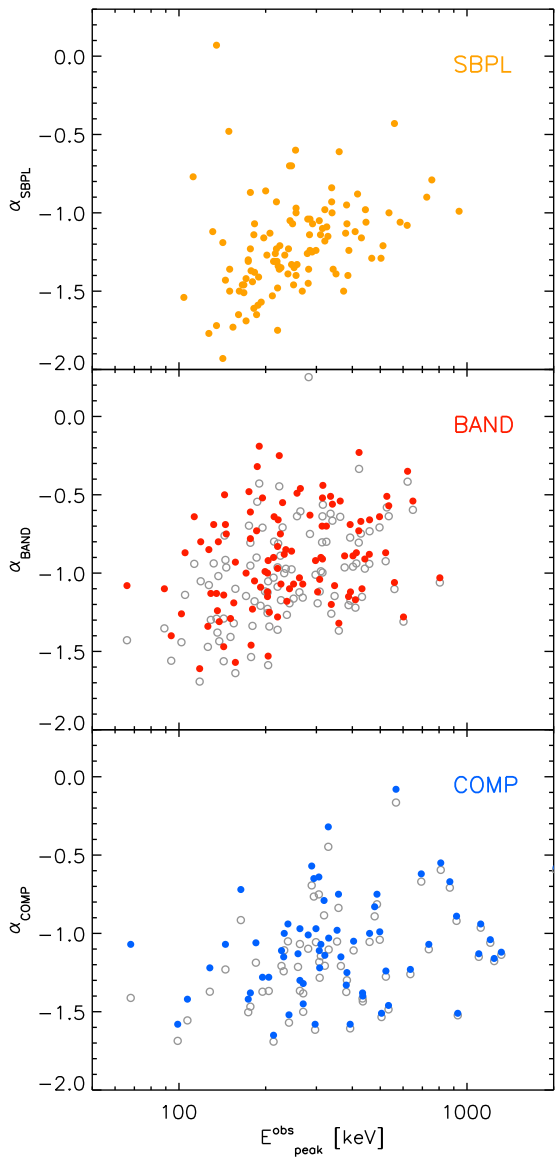}
\caption{
Bright BATSE bursts: low energy power law index vs \epo\ for different best fit 
models (SBPL, BAND and COMP, respectively). 
In the middle and bottom panels, filled circles represent $\alpha$ derived from 
the best fit and empty ones $\alpha_{\rm eff}$ estimated using Eq. \ref{aeff}.} 
\label{alfa_ep_k06} 
\end{figure} 

\citet{preece98a} and \citet{preece02} derive the low energy power--law 
index of a sample of 156 time 
integrated and 5500 time resolved spectra of BATSE bursts 
and discuss its inconsistency with the predicted properties  of the simplest 
synchrotron emission models (see also \citealt{ghirlanda03}).

However they do not consider the values of  $\alpha$ derived 
directly from the fit but the effective 
low energy spectral index $\alpha_{\rm eff}$, defined by \citealt{preece98a} (also adopted by K06)
as
\begin{equation}
\alpha_{\rm eff}=\alpha-(\alpha+2)\frac{E_{\rm fid}}{E^{\rm obs}_{\rm peak}},
\label{aeff}
\end{equation}
which represents the effective slope of the  fitted spectral model at a fiducial energy 
$E_{\rm fid}$. For the BATSE instrument they set $E_{\rm fid}=25$ keV. 

In the case of the SBPL model the introduction of $\alpha_{\rm eff}$ is not 
necessary, since the low energy index retrieved from the fit already describes 
the spectral slope at low energies.

For the BAND and the COMP models, instead, $\alpha$ represents an asymptotic value and 
is always harder than $\alpha_{\rm eff}$. The difference between the two slopes 
depends (for a fixed value of $E_{\rm fid}$) only on \epo. 
For high \epo, $\alpha$ and $\alpha_{\rm eff}$ are very similar, since 
the asymptotic $\alpha$ well describes the true slope of the data 
in the low energy part of the spectrum.
When \epo\ is close to $E_{\rm fid}$=25 keV, 
$\alpha$ and $\alpha_{\rm eff}$ differ significantly, since $\alpha_{\rm eff}$ is 
strongly affected by the spectral curvature.

In Fig. \ref{alfa_ep_k06} we show, for all the three spectral models, 
$\alpha$ as a function of \epo\ (filled circles). 
For the BAND and COMP models, also $\alpha_{\rm eff}$ vs \epo\ (empty circles) is plotted.
As expected, $\alpha_{\rm eff}$ is on average softer than 
$\alpha$ and the difference is more evident for low values of \epo. 
Note that this difference is indeed less pronounced 
for bursts best fit with a
COMP model as this returns on average larger \epo\ (see \S 4.1).
In the case of the SBPL model there is a clear trend between $\alpha$ 
and \epo, revealing that the value of $\alpha$ is strongly affected 
by the energy where the spectrum curves with respect to the low energy bound
of sensitivity of the instrument.
A trend between $\alpha$ and \epo\ is weakly 
present also for burst best fit by the BAND model, and its significance increases when 
considering $\alpha_{\rm eff}$ in place of $\alpha$.

Given these systematic effects introduced by the definition of 
$\alpha_{\rm eff}$ and the arbitrary choice of $E_{\rm fid}$ 
which is related to the energy range of the instrument, 
we rely on the fitted $\alpha$ values
for a comparison between BATSE and GBM bursts.  

At first we consider the BATSE spectra of K06 for the three models (upper panel of
Fig. \ref{a_fermi}).  The corresponding central values of the gaussian fits are reported in 
Table \ref{modelli}. 
We note that for the effects discussed above the $\alpha$ distribution 
for the COMP model is softer than for the BAND one  
(KS probability $P=2\times10^{-4}$) and that the SBPL one provides 
the softest $\alpha$ distribution  (KS probability $P=5\times10^{-9}$ 
compared with the BAND and $P=$ 0.046 with the COMP ones). 
This is also the case when $\alpha_{\rm eff}$ is considered 
(see $\alpha_{\rm eff,25}$ in Table \ref{modelli}). 

Note that the fact that the fitted values of $\alpha$ are harder than 
$\alpha_{\rm eff}$ makes the inconsistency of GRB spectra with the 
limits of synchrotron theory even stronger \citep{preece98a,preece02,ghisellini00,ghirlanda03}.

The bottom panel of Fig. \ref{a_fermi} shows the $\alpha$ distribution for 
\fe/GBM bursts, fitted with either the BAND or the COMP model. 
The central values of the two distributions are reported in Tab. \ref{modelli}. 
Also for \fe\ bursts we find that the COMP model provides values of $\alpha$ 
softer than the BAND one,  but in this case 
the difference is less significant (KS probability $P=0.017$). 
Interestingly the $\alpha$ distribution of 
\fe\ bursts (for both models) is harder than that of BATSE bursts:
the KS probability that the two 
belongs to the same parent population of is 
$P = 6\times10^{-9}$ (last row in Table \ref{modelli}). This probability increases when 
excluding from the sample of BATSE bursts those fitted with a SBPL model 
($P=0.005$ -- next to last row). For comparison, we estimate the effective low energy index at $E_{\rm fid} = 25$ keV
$\alpha_{\rm eff,25}$ 
(Table \ref{modelli}). Also in this case no matter the model its value is systematically 
harder compared to $\alpha_{\rm eff,25}$ estimated for BATSE bursts. For the 
GBM we also report $\alpha_{\rm eff,8}$, i.e. 
estimated at the low energy bound of the GBM instrument 
(8 keV). For this instrument the difference between $\alpha$ and 
$\alpha_{\rm eff,8}$ is not statistically significant ($P = 0.26$ when considering 
the whole GBM sample). We conclude that for the GBM the definition 
of $\alpha_{\rm eff}$ is not meaningful, as the value of $\alpha$ retrieved from the 
fit is an accurate estimate of the true slope at the low energy bound 
of the spectrum.
\begin{figure}
\includegraphics[scale=0.55]{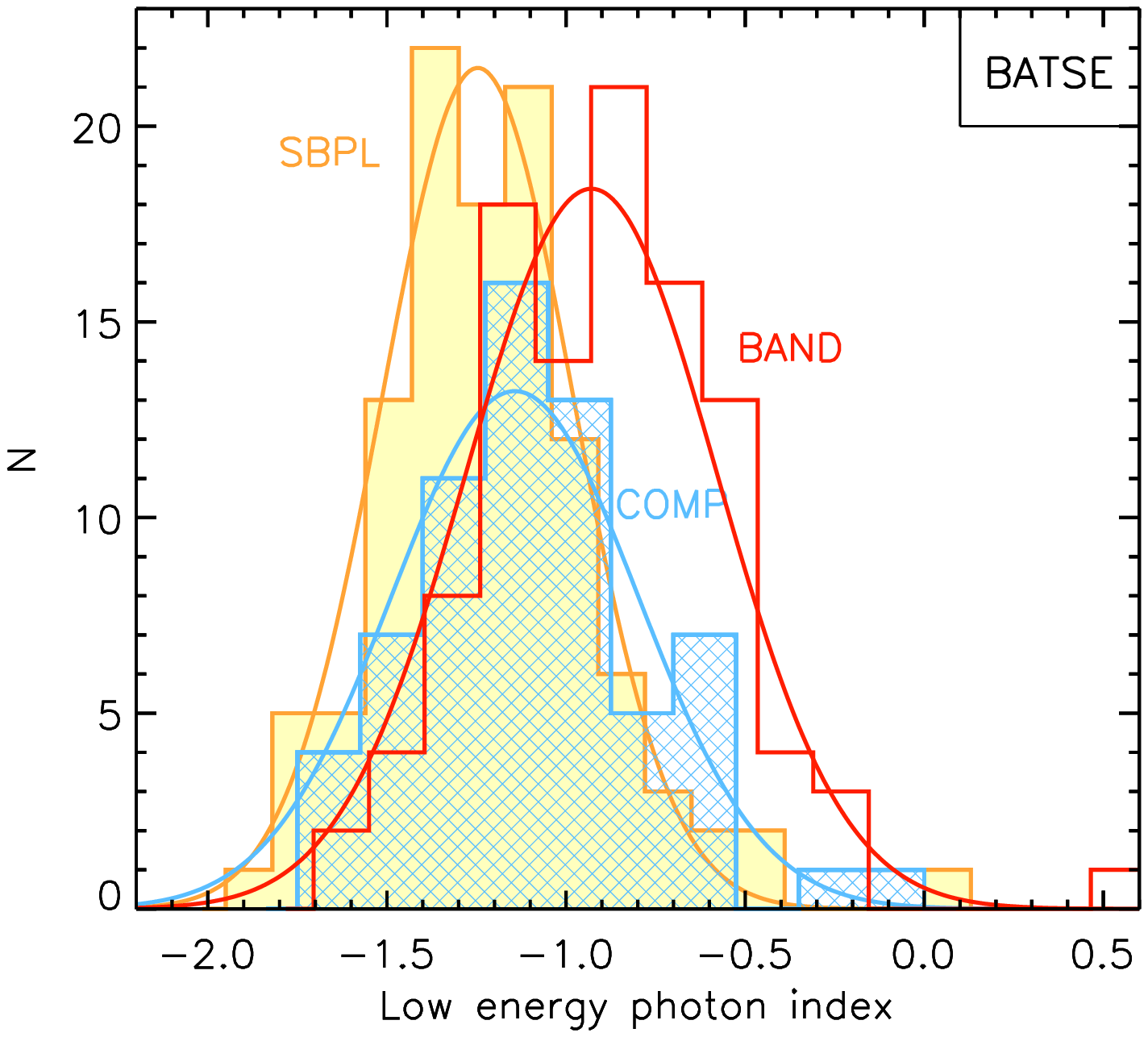}
\includegraphics[scale=0.55]{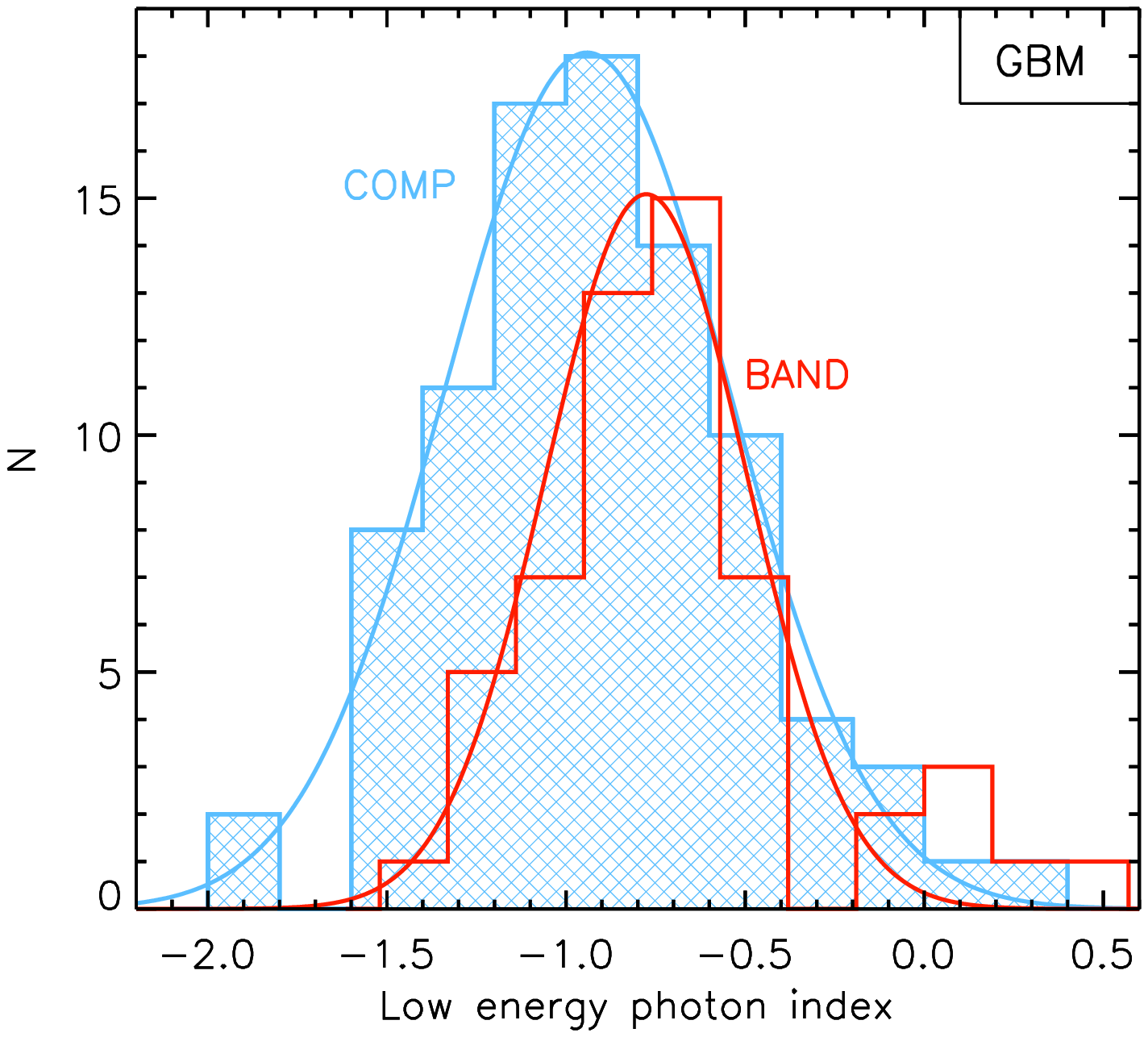}
\caption{
Comparison of the low energy spectral index distributions for the 
according to the spectral best fit model for: BATSE bursts (from K06, upper panel) and for the \fe/GBM 
bursts (this work, bottom panel). For the latter the  solid and hatched histograms
correspond to the BAND and COMP model, respectively. } 
\label{a_fermi} 
\end{figure} 

\section{Discussion and Conclusions}
\begin{figure*}
\vskip -2 true cm
\includegraphics[scale=0.6]{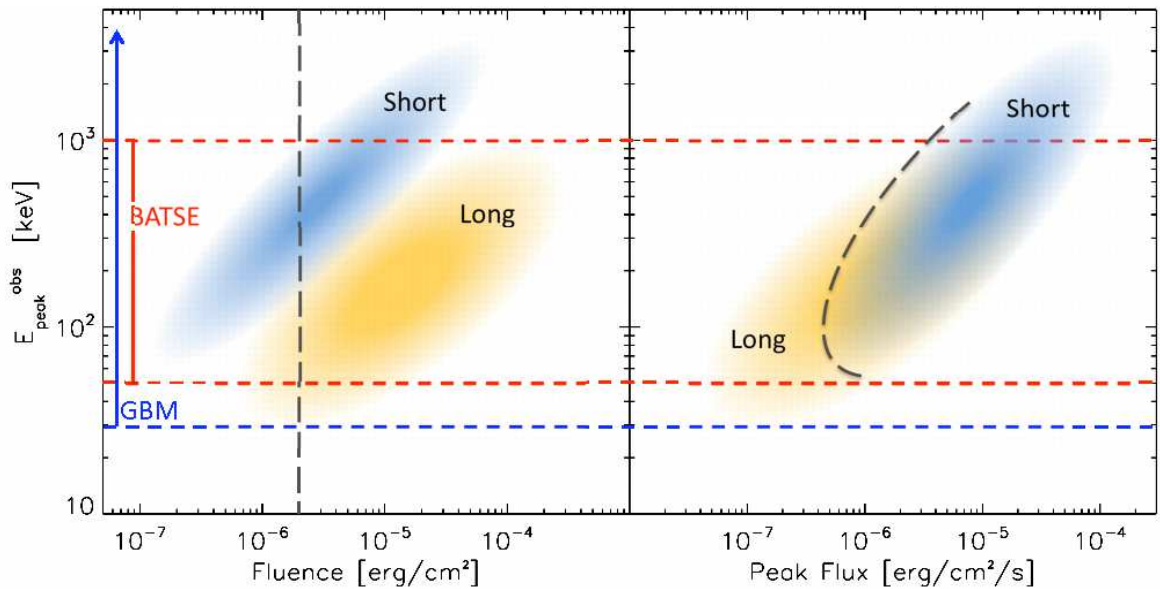}
\vskip -2 true cm
\caption{
Schematic view of the distribution of long and short GRBs in the \epof\ 
and \epop\ planes. The horizontal dashed line at $\sim$ 30 keV represents 
the lower limit for the GBM instrument: the simulations performed in this work 
show that \epo\ can be hardly determined below this value. 
For the BATSE instrument this limit corresponds to $\sim$ 
50 keV. 
The upper limit for BATSE is at $\sim$ 1 MeV, while for the GBM there is no 
upper limit in this plane. The vertical dashed line (left 
panel) shows an example of fluence selection, while the dashed curve (right 
panel) refers to the photon flux selection criterion adopted by G09.}

\label{figurella} 
\end{figure*} 

In this work we study the spectral properties of \fe\
bursts detected by the GBM instrument up to February 2010. 
We collect a sample of 227 GRBs (37 of which 
are classified as short) from the GCN circulars. 
For 146 long and 20 short bursts the spectrum could be fitted by a 
curved model so that the peak energy of the $\nu F_{\nu}$ spectrum 
could be constrained. 

The population of long bursts is large enough to allow a statistical comparison 
with the BATSE results. We consider a complete sample of BATSE bursts 
down to a limiting fluence of $F\sim10^{-6}$ erg cm$^{-2}$. 
Fig. \ref{fluence_tutti} shows that 
the two samples have very similar fluence distributions. 
For both instruments the \epo\ 
distribution peaks around \epo$\sim 150$ keV (Fig. \ref{ep_tutti}). Despite of 
its larger energy range, the GBM extends the \epo\ distribution 
of long bursts 
only at low energies with respect to BATSE. The $\alpha$ distribution, 
instead, reveals some difference: GBM bursts have on average a harder 
low energy photon index ($\langle\alpha_{\rm GBM}\rangle=-0.9$ and 
$\langle\alpha_{\rm BATSE}\rangle=-1.1$, KS probability $P=6\times10^{-9}$). 
Also for GBM short bursts we can 
draw some conclusions about their spectral properties. Their \epo\ distribution is 
shifted towards higher energies  compared both to long bursts from the same 
instrument and to short bursts seen by BATSE. The lack of low energy \epo\ 
(below $\sim 250$ keV) can be accounted for by the spectral 
threshold we derived for the GBM instrument (see Fig. \ref{piani correle}), while 
the larger energy 
coverage allows the detection of \epo\ up to $\sim$ 4 MeV. 
\fe\ data confirm that short bursts have on average a harder $\alpha$ 
compared to long bursts ($\langle\alpha_{\rm GBM,short}\rangle\sim -0.46$),
as already found in the BATSE sample by G09.

Similarly to what has been done for long and short GRBs detected by 
BATSE (N08; G09) we examine the distribution of \fe\ bursts in the 
\epo--Fluence and \epo--Peak Flux planes in order to study instrumental selection effects 
and test their consistency with the 
rest frame correlations (i.e. \ama\ and \yone, respectively) defined by 
GRBs with measured redshifts.  

Our main results are:
\begin{itemize}

\item {\it Long GRBs} detected by \fe\ follow the same \epof\ and \epop\ 
correlations defined by BATSE GRBs (Fig. \ref{piani correle}). 
We computed the instrumental selection effects 
of \fe/GBM -- as already done for BATSE (G08; N08): 
the trigger threshold and the spectral 
threshold are not responsible for the correlations 
defined by long GRBs in both planes
(see Fig. \ref{piani correle}).
The \fe/GBM spectral range down to 8 keV 
with respect to the limit of 30 keV of BATSE,  allows to 
extend the correlations to lower peak energies/fluences.
Instead, despite of the higher energy sensitivity of 
\fe/GBM (up to 40 MeV) no long GRBs with \epo\ larger than a few MeV are
detected. 
This can be due to a real absence of bursts
with such high \epo\ or to the fact that they have large fluences, 
thus being too rare to be detected during 1.5 year of \fe\ observations.

We conclude that long GRBs detected by \fe\ confirm what 
found with BATSE bursts, in particular that they follow a correlation both 
in the \epof\ and in the \epop\ plane. 
Moreover, the fraction of bursts detected by \fe\ which are outliers at more 
than 3$\sigma$ with respect to the \ama\ correlation is $\sim$5\%, similarly  
to what found in the BATSE sample, 
while there are no outliers (at more than 3$\sigma$) of the \yone\ correlation 
among \fe\ long GRBs.

\item {\it Short GRBs} detected by \fe\ populate a different region in 
the \epof\ plane with respect to long events, the firmer having a larger peak 
energy and lower fluences compared to the latter ones. 
This is consistent with what found by BATSE 
and confirms  that short GRBs do not follow 
the "Amati" correlation but they obey the "Yonetoku" correlation 
defined by long events.
\end{itemize}

The comparison of short and a representative sample of long BATSE GRBs 
(selected with a similar peak flux threshold) led G09 to conclude 
that their main spectral diversity is due to a 
harder low energy spectral index in short bursts while their 
\epo\ of BATSE is similarly distributed. 
 \fe\ bursts provide the 
the opportunity of re-examining this result for the population of 
short and long GRBs detected by the GBM and also compare their spectral 
properties with those of the  BATSE ones. We find that:

\begin{itemize}
\item 
\epo\ of short \fe\ GRBs is larger and \al\ smaller that those of long one, indicating that 
short events are harder, both in terms of their peak energy and low energy spectral index. 

\item 
A comparison between GBM and BATSE short bursts reveals that they 
have similar \al\ while the \epo\ of short GBM bursts is larger 
than that of short BATSE events (see Fig. \ref{piani correle}, bottom left panel). This 
information is allowed by the higher energies which can be detected by the GBM.
Moreover, the different \epo\ distribution of BATSE and GBM short 
bursts is affected by the lower sensitivity of the GBM instrument, which 
misses short bursts at low fluence (and therefore low \epo).

\item 
\fe\ and BATSE long bursts have a similar \epo\  while 
\fe\ events tend to have a harder low energy spectral 
index (Fig. \ref{a_fermi}).

\end{itemize}

Fig. \ref{figurella} shows a schematic representation of the current 
information about the distribution of short and long bursts in the \epof\ and 
\epop\ planes. With respect to BATSE, the GBM reveals that long bursts 
extend to lower \epo, consistently with what previously found with other 
instruments (mainly Hete--II and Swift). 

Despite of the high energy sensitivity, also the \epo\ distribution of \fe\ long events
extends only up to $\sim$ 1 MeV. The situation is different for short GRBs
whose \epo\ reach up to $\sim$ 4 MeV in the present sample. 
These high \epo\ were  
not detectable by BATSE, whose sensitivity drops at $\sim$ 1 MeV (upper 
horizontal dashed line in Fig. \ref{figurella}). Therefore, 
the \fe/GBM shows that short GRBs have larger \epo\ with respect to long ones, 
contrary to what found with BATSE (G09). 

When comparing the \epo\ distribution of short and long bursts, different 
conclusions can be drawn, according to the selection criterion of the samples. 
The left panel in Fig. \ref{figurella} shows that 
a given cut in fluence (represented 
by the vertical dashed line) would result in different \epo\ distributions between short 
and long bursts, resulting from their different location in the \epof\ plane. The right panel 
in Fig. \ref{figurella} illustrates, instead, what happens for a selection in {\it photon flux}. This 
translates into a curve in {\it energy 
flux}: the dashed curve represents the cut 
applied by G09 to select both short and long bursts, corresponding to 
a photon flux larger than 3 ph cm$^{-2}$s$^{-1}$. 
This criterion applied to BATSE bursts 
produces similar \epo\ distributions of long and short events, as indeed found by 
G09. The very same criterion applied to \fe/GBM bursts results, instead, in different 
distributions, since short bursts can have very 
high \epo\ values, not detected in the sample of long bursts.

\begin{acknowledgements} 
We acknowledge the GBM team for the public distribution of the spectral properties of \fe/GBM bursts through the GCN network. 
We also thank Y. Kaneko for private communications on the BATSE bright sample spectral results. 
This research has made use of data obtained through the High Energy Astrophysics Science Archive 
Research Center Online Service, provided by the NASA/Goddard Space Flight Center. 
This work has been partly supported by ASI grant I/088/06/0. LN thanks the Osservatorio Astronomico di Brera for the 
kind hospitality for the completion of this work. 
\end{acknowledgements}
\bibliographystyle{aa} 
\bibliography{biblio}
\end{document}